\documentclass[letterpaper, 10 pt, conference]{ieeeconf}  

\IEEEoverridecommandlockouts                              

\usepackage{graphics} 
\usepackage{epsfig} 
\usepackage{amsmath,amssymb,mathtools}
\usepackage{graphicx}

\usepackage[table]{xcolor}
\usepackage{booktabs}
\usepackage{newtxtext}
\usepackage{newtxmath}        
\usepackage[cal=cm]{mathalfa} 

\usepackage{amsmath} 
\usepackage{amssymb}  
\usepackage{soul,color}

\usepackage{amsmath,amssymb,amsfonts,amstext}
\usepackage{graphicx}
\usepackage{textcomp}
\usepackage{babel}
\usepackage{balance}
\usepackage{comment}
\usepackage{todonotes}
\usepackage{mathtools}

\usepackage{algorithm}
\usepackage{algorithmicx} 
\usepackage{algpseudocode}

\makeatletter
\let\NAT@parse\undefined
\makeatother

\usepackage{tikz}
\usetikzlibrary{shapes, arrows.meta, positioning}
\usepackage{amsmath}
\usepackage{float} 
\usepackage{graphicx}
\usepackage{caption}
\usepackage{subcaption}
\usepackage{dblfloatfix}

\title{\LARGE \bf
Spatiotemporal Forecasting of Incidents and Congestion with Implications for Sustainable Traffic Control
}

\author{Tony Kinchen, Ting Bai, Nishanth Venkatesh S., and Andreas A. Malikopoulos
\thanks{This research was supported in part by NSF under Grants CNS-2401007, CMMI-2348381, IIS-2415478, and in part by MathWorks.}\vspace{1.5pt}
\thanks{T. Kinchen and N. V. Senthil Kumar are with the Systems Engineering program, Cornell University, Ithaca, NY, USA. Email: \{{\tt\small tjk238, ns942\}@cornell.edu}}
\thanks{T. Bai and A. A. Malikopoulos are with the  School of Civil $\&$ Environmental Engineering, Cornell University, Ithaca, NY, USA. E-mails: \{{\tt\small tingbai, amaliko\}@cornell.edu}}}

\begin{document}
\maketitle

\begin{abstract}
Urban traffic anomalies, such as collisions and disruptions, threaten the safety, efficiency, and sustainability of transportation systems. In this paper, we present a simulation-based framework for modeling, detecting, and predicting such anomalies in urban networks. Using the Simulation of Urban MObility (SUMO) platform, we generate reproducible rear-end and intersection crash scenarios with matched baselines, enabling controlled experimentation and comparative evaluation. We record vehicle-level travel time, speed, and emissions for both edge- and network-level analysis. Building on this dataset, we develop a hybrid forecasting architecture that combines bidirectional long short-term memory networks with a diffusion convolutional recurrent neural network to capture temporal dynamics and spatial dependencies. Our simulation studies on the Broadway corridor in New York City demonstrate the framework's ability to reproduce consistent incident conditions, quantify their effects, and provide accurate multi-horizon traffic forecasts. Our results highlight the value of combining controlled anomaly generation with deep predictive models to support reproducible evaluation and sustainable traffic management.
\end{abstract}

\section{Introduction}\label{Section I}
Traffic congestion and unexpected incidents, such as collisions, are recurring phenomena in urban transportation systems. These events disturb normal traffic flow, increase travel times, and exacerbate fuel consumption and emissions, leading to negative impacts on urban mobility~\cite{choudhary2022urban,11301032}. Unlike other disruptions, congestion induces stop-and-go driving patterns, where prolonged idling and frequent acceleration cycles raise carbon emissions (CE) sharply, especially when average freeway speeds fall below 45-50 mph (72-80 km/h)~\cite{barth2008real}. Mitigating congestion is therefore critical not only for improving traffic efficiency but also for reducing environmental impacts. Despite advances in vehicle platooning~\cite{mahbub2023_automatica,10209062,9993403}, optimal route planning~\cite{le2024stochastic,Malikopoulos2019CDC} and recommendation~\cite{bang2021AEMoD,11312449}, and intersection signal control~\cite{malikopoulos2019ACC,Malikopoulos2020}, traffic congestion remains an inevitable challenge. As urban road networks grow denser and more interconnected, the need for effective approaches to detect and predict traffic anomalies has become increasingly urgent to enhance system resilience and promote sustainable mobility. 

Conventional incident detection approaches typically rely on predefined thresholds or rule-based models that compare observed traffic parameters (e.g., vehicle speed, occupancy, traffic flow) against nominal ranges, triggering alarms when significant deviations are detected~\cite{elsahly2022systematic}, \cite{stephanedes1992comparative}. While these methods provide foundational baselines, they fall short in capturing the nonlinear and evolving dynamics of urban traffic systems, particularly in traffic environments where high variability and spatiotemporal dependencies often undermine detection accuracy.

Recent research has explored data-driven techniques for identifying anomalies in traffic systems. Graph-based models and recurrent neural networks have been employed to forecast traffic states and identify irregularities in congestion patterns~\cite{xu2019anomaly}. In parallel, predictive modeling has advanced with techniques such as sequence-to-sequence learning and diffusion-based recurrent frameworks, which are widely adopted for spatiotemporal forecasting~\cite{zhao2021prediction}. Attention-based mechanisms further extend these approaches by improving long-horizon prediction accuracy and reducing sensitivity to noise in sensor data. Informer-inspired architectures and Siamese network designs have also shown promise in capturing long-term dependencies and distinguishing abnormal traffic conditions~\cite{peng2022traffic, li2020deepflow}. More recently, informer-based models applied to urban mobility~\cite{kang2023traffic} have demonstrated robust performance in anomaly detection tasks, achieving higher recall and F1 (a metric used to evaluate model accuracy) scores than competing baseline methods.  

Video-based methods provide another important avenue for traffic anomaly detection. Deep learning pipelines that integrate convolutional neural networks, long short-term memory networks (LSTMs), and fuzzy reasoning have been developed to capture spatiotemporal cues for accidents and near-misses in traffic streams~\cite{khan2022anomaly}. Rough–fuzzy models and temporal attention mechanisms further extend these capabilities to structured video summarization and anomaly classification~\cite{pramanik2022traffic,natha2025deep}. Moreover, edge–server architectures have been introduced to support the detection of abnormal pedestrian behaviors in large-scale closed-circuit television environments, enabling scalable and real-time analysis~\cite{song2024pedestrian}.

In this paper, we propose a simulation-based framework that unifies anomaly generation, metric logging, and predictive modeling. Using the SUMO platform, we construct reproducible collision scenarios and matched baselines for systematic evaluation. The framework records fine-grained per-vehicle metrics, enabling both localized and network-wide analysis, including travel time index comparisons~\cite{erdelic2021congestion}. We then develop a hybrid forecasting architecture combining bidirectional LSTM (BiLSTM) networks with diffusion convolutional recurrent neural networks (DCRNN) to capture both temporal dependencies and spatial dynamics~\cite{li2018dcrnn}. This approach enables real-time prediction of traffic disruptions and reveals how controlled anomalies impact urban mobility. Note that the BiLSTM--DCRNN architecture follows established spatiotemporal forecasting practice and is adopted here as a testbed rather than as a methodological contribution. The main contributions of this paper are as follows: (i) We design a reproducible SUMO-based incident generation pipeline with matched baseline scenarios that enables controlled and fair benchmarking.
(ii) We establish a unified metric collection and temporal–spatial alignment protocol for vehicle-, edge-, and network-level analysis.
(iii) We develop an interval-based spatiotemporal localization and forecasting formulation that links event-level collision containment with post-incident traffic propagation. Simulation results demonstrate high spatiotemporal containment accuracy for collision localization and accurate multi-horizon forecasts of congestion and emissions, showing the validity of the proposed integrated framework.


\section{Problem Formulation}\label{Section II}
We consider a mixed urban traffic network simulated in SUMO with scripted collision events. The road network is characterized as a directed graph $\mathcal{G}(\mathcal{N}, \mathcal{E})$, where $\mathcal{N}$ is the node-set with each node representing a road intersection, an entry, or an exit point. Edges in $\mathcal{E}$ denote road segments connecting pairs of nodes. Time is discretized as $\mathcal{T}\!=\!\{1,\dots,T\}$. For each edge $e\!\in\!\mathcal{E}$ and time $t\!\in\!\mathcal{T}$, let $f_{e,t}\!\in\!\mathbb{R}^2$ denote a feature vector comprising the travel time index (TTI) and CE, which capture the congestion level and environmental cost on that edge, respectively. More specifically, we define the TTI~\cite{erdelic2021congestion} on edge $e$ at time $t$ as $\mathrm{TTI}_{e,t} = \mathrm{TT}_{e,t}/\mathrm{TT}^{\mathrm{ff}}_{e}$, where $\mathrm{TT}_{e,t}$ is the observed travel time to traverse edge $e$ based on the real traffic flow at time $t$, and $\mathrm{TT}_e^{\mathrm{ff}}$ denotes the free-flow travel time without congestion. CE captures all vehicles' emissions on edge $e$ within the time interval $[t-1,t]$, and we describe it using a fitted polynomial function of vehicle speed $v$ (m/s) and acceleration $a$ (m/s$^2$), expressed as $CE(v, a) = c_0 + c_1 v a + c_2 v a^2 + c_3 v + c_4 v^2 + c_5 v^3$, where the coefficients $c_0, \dots, c_5$ are determined empirically for each pollutant and vehicle class~\cite{Krajzewicz2014}. These two metrics, TTI and CE, form the basis for modeling network-level traffic states.

With these edge-level definitions of travel time and emissions, we next specify the vehicle classes and admissible events used in our scenarios. The vehicle set $\mathcal{V}$ comprises passenger vehicles (PVs), buses, and automated vehicles (AVs), i.e., $\mathcal{V}\!=\!\{\text{PV}, \text{bus}, \text{AV}\}$. Event admissibility follows \(\mathcal{R}\), which we define as
\begin{align}
\mathcal{R} &:= \big\{
   (\text{PV}, \text{PV}, \text{rear}), \ 
   (\text{PV}, \text{bus}, \text{rear}), \ (\text{PV}, \text{AV}, \text{rear}),  \nonumber\\
   &  \ \ \ \ \ \ \ \  \
   (\text{PV}, \text{PV}, \text{intersection})
   \big\},
\end{align}
where the second vehicle in each combination is the leading vehicle in the interaction. For intersections, we treat the scenario under the assumption that PVs exhibit more reckless driving and uncoordinated entry into unsignalized intersections, unlike buses with professional drivers or AVs with automated stopping features, and therefore restrict events to PV-PV interactions. In this paper, we aim to present a framework that is capable of reproducing consistent incident conditions, quantifying their effects, and providing accurate multi-horizon traffic forecasts. To this end, we employ a spatiotemporal forecasting model, DCRNN, for predicting congestion and emissions, complemented by a BiLSTM module that provides interval estimates of collision time and location.

\subsection{Collision Scripting and Admissibility}
We use collision scripting to introduce controlled crashes into the SUMO simulation to reproduce repeatable disruption scenarios. Let $\mathcal{V}\!=\!\{\text{PV}, \text{bus}, \text{AV}\}$ denote the set of vehicle classes, $\mathcal{S}\!=\!\{\text{rear}, \text{inter}\}$ the set of event types, and $\mathcal{R}\!\subseteq\! \mathcal{V}\!\times\!\mathcal{V}\!\times\!\mathcal{S}$ the admissibility relation. Only events that lie in $\mathcal{R}$ are retained for learning and evaluation. Specifically, we consider $(\text{PV},\text{PV},\text{rear})$, $(\text{PV},\text{bus},\text{rear})$, $(\text{PV},\text{AV},\text{rear})$, and $(\text{PV},\text{PV},\text{inter})$. Two event classes are scripted: rear-end and intersection. We assume rear-end events to occur among PV-PV, PV-bus, and PV-AV, while intersection conflicts are restricted to PV-PV. Events are inserted by halting vehicles at designated locations for a dwell duration and resuming movement thereafter; logs record $(x,y,t)$, where $x,y\!\in\!\mathbb{R}$ are the spatial coordinates of the collision in meters, and $t\!\in\!\mathcal{T}$ denotes the simulation time in seconds.

\subsection{Spatiotemporal Forecasting}
We aim to forecast edge-level values of TTI and CE over multiple horizons. Let $f_{e,t}$ denote the feature vector for edge $e\!\in\!\mathcal{E}$ at time $t\!\in\!\mathcal{T}$. The forecasting problem is defined as learning to approximate the true target values $z^{(q)}_{e,t}$, where $z^{(q)}_{e,t}$ represents the scalar corresponding to metric $q\!\in\!\{\mathrm{TTI}, \mathrm{CE}\}$ from past observations. For a forecast origin $\tau$, the model ingests the recent lookback window $\{f_{e,\tau-L+1},\dots,f_{e,\tau}\}$ across all edges, where $L$ is the look-back length and $|\mathcal{E}|$ is the number of edges. The model then produces forecasts $\{\hat{z}^{(q)}_{e,\tau+h}\}_{h=1}^H$, where $H$ is the forecast horizon and each $\hat{z}$ corresponds to a prediction of TTI or CE.

The training objective reflects the interval nature of predictions. Let $\beta$ be a tunable hyperparameter that controls the trade-off between interval sharpness and coverage. For each target quantity $q\!\in\!\{\mathrm{TTI}, \mathrm{CE}\}$, the model outputs lower and upper bounds $(\widehat{z}^{(q)}_{\text{low}}, \widehat{z}^{(q)}_{\text{high}})$. The loss penalizes under- and over-coverage while encouraging tight intervals, with additional weighting for rare high-magnitude spikes by
\begin{align}
\mathcal{L}\!=\! 
\frac{1}{N}\!\sum_{i=1}^N\!\Big[
&\beta\big(\widehat{z}^{(q)}_{\mathrm{high},i}\!-\!\widehat{z}^{(q)}_{\mathrm{low},i}\big)
+\big(\max\{0,\widehat{z}^{(q)}_{\mathrm{low},i}\!-\!z^{(q)}_i\}\nonumber \\
&+ \max\{0,z^{(q)}_i\!-\!\widehat{z}^{(q)}_{\mathrm{high},i}\}\big)\!\cdot\!w^{(q)}_i
\Big],
\end{align}
where $N$ is the number of training samples, $w^{(q)}_i$ is a target-specific weight that emphasizes rare spike events in either TTI or CE, and $\beta\!>\!0$ penalizes excessively wide prediction intervals. At inference time, intervals are further calibrated using where residuals on a held-out calibration set determine edge-specific padding values to ensure desired coverage.

\subsection{Event Localization with BiLSTM}
To localize collision occurrences, our objective is to predict when and where an event will take place in terms of its spatial coordinates $(x,y)$ and time $t$. For each dimension $z\!\in\!\{x,y,t\}$, the model utilizes the upper and lower predictions to define an interval expected to contain the true event value $z^\ast$. We train the BiLSTM regressor by minimizing the mean squared error between the predicted bounds and the observed values, encouraging the intervals to be both accurate and appropriately tight. We evaluate model performance using root mean squared error of the endpoints, the probability that true events fall inside the predicted intervals (coverage), the average width of the predicted ranges, and the Dice score, which quantifies overlap between predicted and actual event intervals.

\section{Modeling Framework}\label{Section III}
This section presents the modeling framework used to capture both the spatiotemporal occurrence of collision events and their impacts on system-level traffic metrics, including TTI and CE. The framework integrates event localization with network-level forecasting to characterize how localized disruptions propagate through urban traffic networks.

\begin{figure}[t]
    \centering
    \includegraphics[width=0.95\columnwidth]{\detokenize{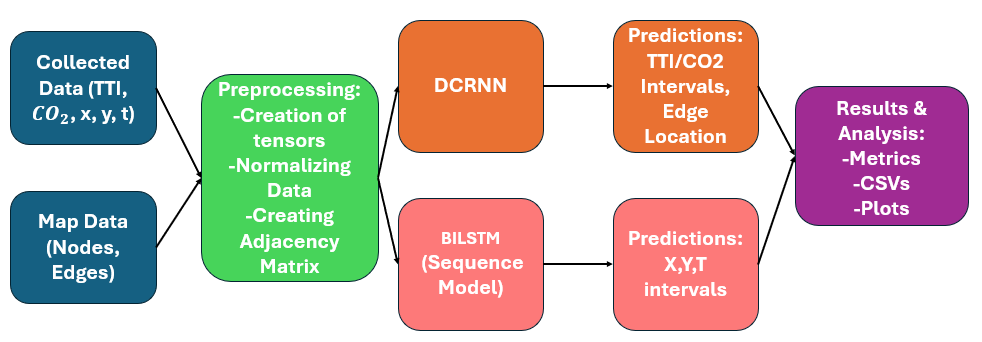}}
    \vspace{-5pt}
    \caption{Flow chart of the modeling framework.}     
    \label{fig:1}
\end{figure}

As illustrated in Fig.~\ref{fig:1}, the framework consists of two complementary components: (i) event localization, which estimates the spatiotemporal coordinates of collision events, and (ii) spatiotemporal forecasting, which predicts the evolution of edge-level traffic states over multiple horizons. Both components rely on recurrent architectures that leverage temporal sequences and graph-structured representations of the road network. Below, we describe the BiLSTM encoder, the diffusion convolution operator, and the diffusion-convolutional recurrent module with fusion layers.

\subsection{BiLSTM Encoder}
Bidirectional long short-term memory (BiLSTM) networks are used to capture temporal dependencies by processing input sequences in both forward and backward directions. Given a sequence of features $\{r^{(t)}\}_{t=1}^{T}$ with $r^{(t)} \in \mathbb{R}^{F_{\mathrm{seq}}}$, the BiLSTM encoder concatenates the final hidden states from both directions to form a representation in $\mathbb{R}^{2d_{\mathrm{LSTM}}}$, where $d_{\mathrm{LSTM}}$ is the hidden dimension in one direction. This encoding captures both past and future temporal context and serves as the basis for spatiotemporal event localization. In implementation, the BiLSTM consists of two stacked bidirectional layers with hidden dimension $d_{\mathrm{LSTM}}$ in each direction. The final forward and backward hidden states are concatenated and passed through three parallel linear output heads that produce lower and upper bounds for each of $x$, $y$, and $t$, yielding a six-dimensional interval representation per prediction horizon.

\subsection{Diffusion Convolution}
To model directed spatial dependencies on the traffic network $\mathcal{G}(\mathcal{N},\mathcal{E})$, we adopt the diffusion convolution introduced by Li et al.~\cite{li2018dcrnn}. The forward and backward row-normalized transition matrices are defined as
\begin{equation}
S_f= D_O^{-1} W, \quad S_b=D_I^{-1} W^\top,
\end{equation}
where $W$ is the weighted adjacency matrix, $D_O=\mathrm{diag}(WM)$, and $D_I=\mathrm{diag}(W^\top M)$, with $M$ encoding network connectivity.


For a graph signal $G\!\in\!\mathbb{R}^{N \times P}$, the $K$-step bidirectional diffusion convolution is given by

\begin{equation}
\sum_{k=0}^{K-1}
\left(\theta_{k,1} S_f^k+\theta_{k,2} S_b^k \right) G_{:,p}, \quad p=1,\dots,P,\nonumber
\end{equation}
where $K$ controls the diffusion depth, $\theta_{k,1}$ and $\theta_{k,2}$ are trainable parameters, and $*_{G}$ denotes diffusion convolution as a weighted polynomial of $S_f$ and $S_b$ applied to $G_{:,p}$. Here, $S_f^k$ and $S_b^k$ denote matrix powers with $S_f^0\!=\!S_b^0\!=\!I$, and $P\!=\!5$ corresponds to the edge-level traffic features. This formulation enables the model to capture both upstream and downstream spatial influences on each edge and has been shown to be effective for multi-horizon traffic forecasting~\cite{sun2023diffusion}. 

\subsection{DCGRU Encoder and Fusion Multi-Layer Perception}
The diffusion-convolutional gated recurrent unit (DCGRU) extends the standard GRU by replacing dense transformations with diffusion convolutions~\cite{li2018dcrnn}. At each step $k=1,\ldots,T_g$, the hidden state $h^{(k)}$ is updated using graph signals $\{G^{(k)}\}$ and the previous hidden state $h^{(k-1)}$, enabling the model to jointly capture temporal evolution and spatial propagation of traffic states. The graph-level representation is obtained by pooling the hidden states
\begin{equation}
g=\mathrm{Pool}(\{h^{(k)}\}_{k=1}^{T_g}) \in \mathbb{R}^{d_{\mathrm{DCGRU}}},
\end{equation}
where $\mathrm{Pool}(\cdot)$ denotes a temporal aggregation function. The pooled representation is concatenated with other encodings and passed through a feed-forward fusion network
\begin{equation}
h = \phi\!\left(W_2\,\phi(W_1 u + c_1) + c_2\right),
\end{equation}
where $\phi(\cdot)$ is a pointwise nonlinearity. Multi-horizon forecasts are produced via step-specific affine heads
\begin{equation}
\hat{p}^{(s)} = U_s h + d_s, \quad s = 1,\ldots,S,
\end{equation}
with $U_s$ and $d_s$ denoting horizon-specific parameters.

\section{Simulation}\label{Section V}

\subsection{Simulation Setup}
\begin{figure}[t]
    \centering
    \includegraphics[width=\columnwidth]{\detokenize{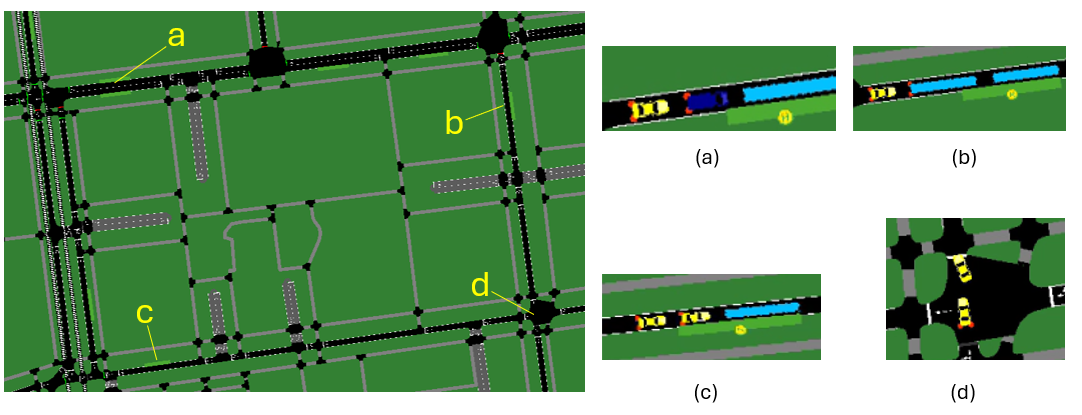}}
    \caption{SUMO map of NYC Broadway intersections. (a) Rear-end (AV-PV). (b) Rear-end (Bus-PV). (c) Rear-end (PV-PV). (d) Intersection (PV-PV).}     
    \label{fig:nyc_intersection}
\end{figure}
As shown in Fig.~\ref{fig:nyc_intersection}, our study focuses on a segment of the Broadway corridor in New York City that includes multiple urban intersections. To capture realistic traffic heterogeneity, we populate the network with PVs, buses, and AVs in proportions summarized in Table~\ref{tab:vehicle_totals}. We applied a mixed routing scenario to reflect diverse travel behaviors across vehicles and to generate realistic interaction patterns for the incident simulations. Training, validation, and test datasets are constructed from independent simulation runs with different random seeds, while preserving identical network geometry, traffic demand, and admissibility rules. This design isolates generalization across stochastic traffic realizations while holding structural conditions fixed, reducing the risk of memorizing deterministic simulator behavior.

\begin{table}[t]
    \centering
    \renewcommand{\arraystretch}{1.} 
    \caption{Vehicle composition and incident involvement}
    \label{tab:vehicle_totals}
    \begin{tabular}{ccc}
        \toprule
        \textbf{~~Vehicle Type~~} & \textbf{~~Total Number~~} & \textbf{~~Incident Involvement~~} \\
        \midrule
        PV   & 330 & Intersection, Rear-end \\
        Bus  & 250 & Rear-end \\
        AV   &  80 & Rear-end \\
        \bottomrule
    \end{tabular}
\end{table}

\subsubsection{Collision Incident}
The simulation process begins with the creation of a collision-injected route file. We use a scripted pre-processing step to insert artificial stopping events into selected vehicle trajectories to simulate two common collision types: rear-end and intersection collisions. A rear-end collision models a PV that does not stop in time when approaching a stationary vehicle ahead, such as a bus at a stop, another PV, or an AV. To capture typical rear-end dynamics in a mixed-traffic network, the lagging PV is assigned the maximum speed, representing a speeding vehicle colliding with the stationary one. We generate intersection collisions as two PVs approaching an intersection from perpendicular directions. One vehicle halts after entering the intersection, while the other approaches in a higher direction from the side, mimicking a scenario where a speeding PV collides with a stopped vehicle inside the intersection. We simulate a collision-injected route file using all vehicles, while omitting collision stops in the control variant. We obtain free-flow baselines by simulating each control variant vehicle independently.

\subsubsection{Congestion Incident}
To quantify the impact of congestion on network performance, we adopt the TTI as described by Erdelic et al.~\cite{erdelic2021congestion}, which serves as a standardized measure of congestion. As introduced earlier in Section II, TTI represents the ratio of actual travel time to an ideal free-flow benchmark, where \(\mathrm{TT}_{e,t}\) is the observed dwell time on edge $e$ during collision-injected simulation and \(\mathrm{TT}^{\mathrm{ff}}_e\) denotes the travel time under free-flow, no-collision conditions. Values greater than $1$ indicate congestion-induced delays. 


An event is considered correctly localized if the true collision coordinates $(x^\ast,y^\ast,t^\ast)$ lie simultaneously within their respective predicted intervals. Figs.~\ref{fig:rearend_t}--\ref{fig:intersection_alt} report marginal containment along individual dimensions for interpretability.


\subsection{Results Analysis}
\subsubsection{Collision Event Localization}
\begin{figure}[t]
    \centering
    \includegraphics[width=0.93\linewidth]{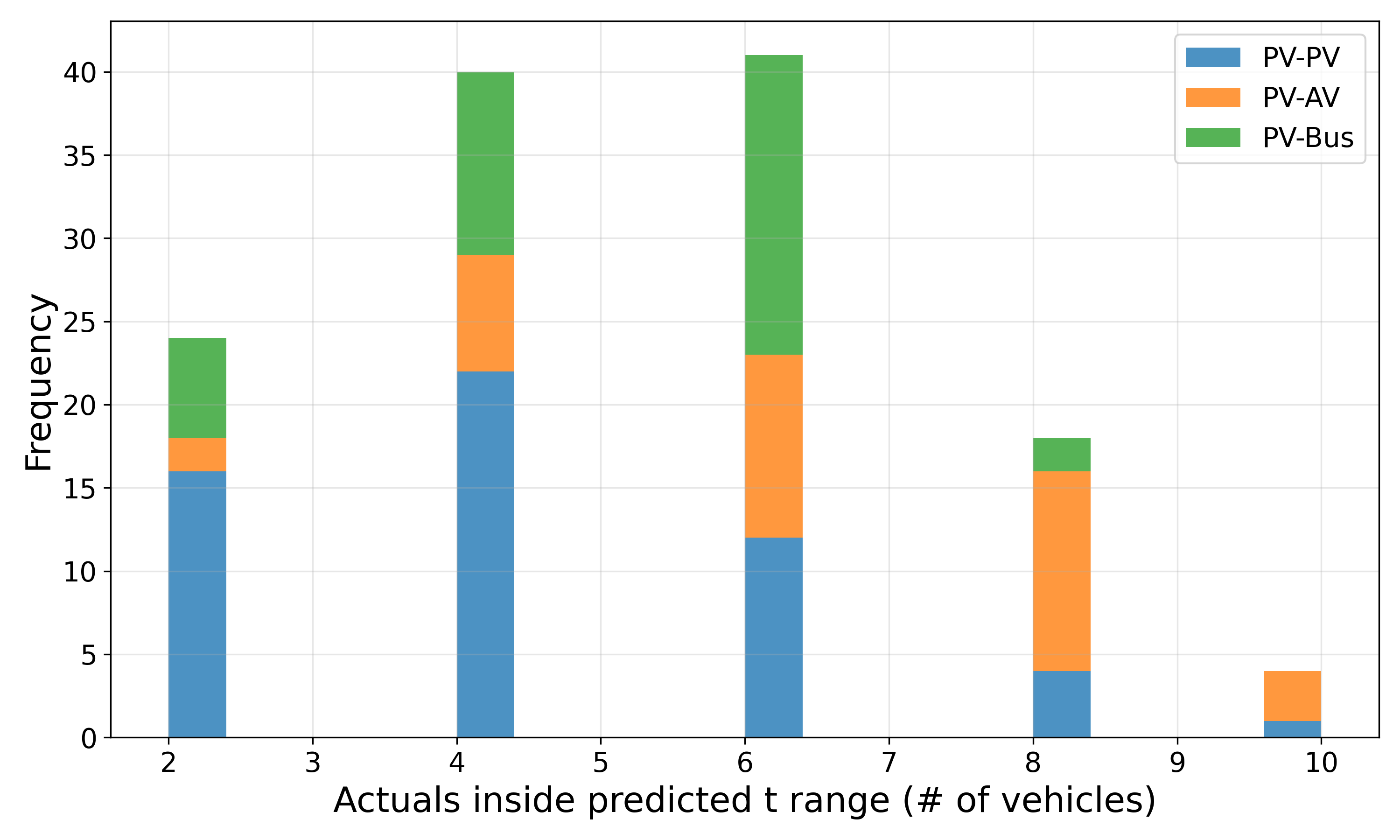}
    \caption{Stacked histogram of actual counts inside predicted $t$ range for rear-end collisions by vehicle pair.}
    \label{fig:rearend_t}
\end{figure}

\begin{figure}[t]
    \centering
    \includegraphics[width=0.93\linewidth]{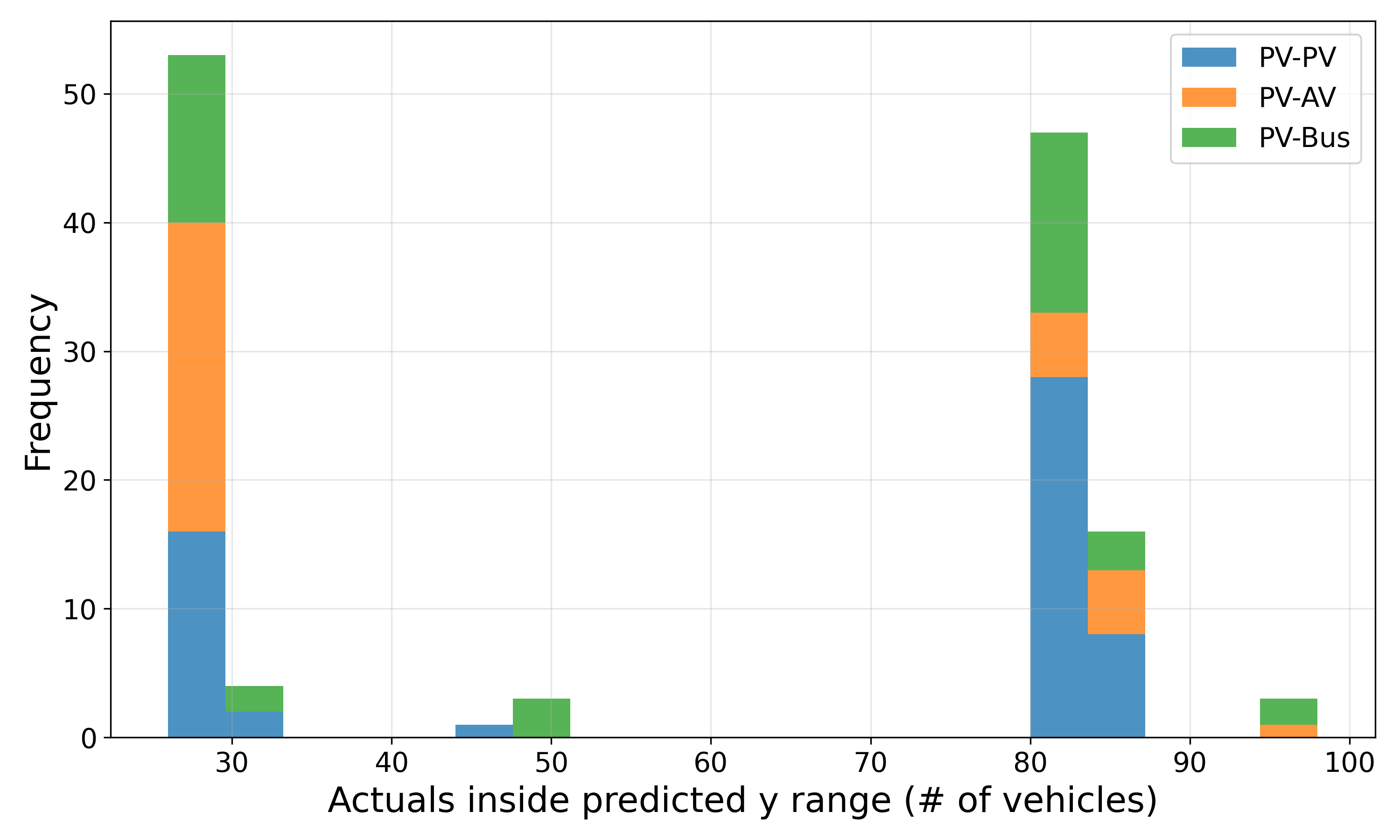}
    \caption{Stacked histogram of actual counts inside predicted $y$ range for rear-end collisions by vehicle pair.}
    \label{fig:rearend_y}
\end{figure}

\begin{figure}[t]
    \centering
    \includegraphics[width=0.93\linewidth]{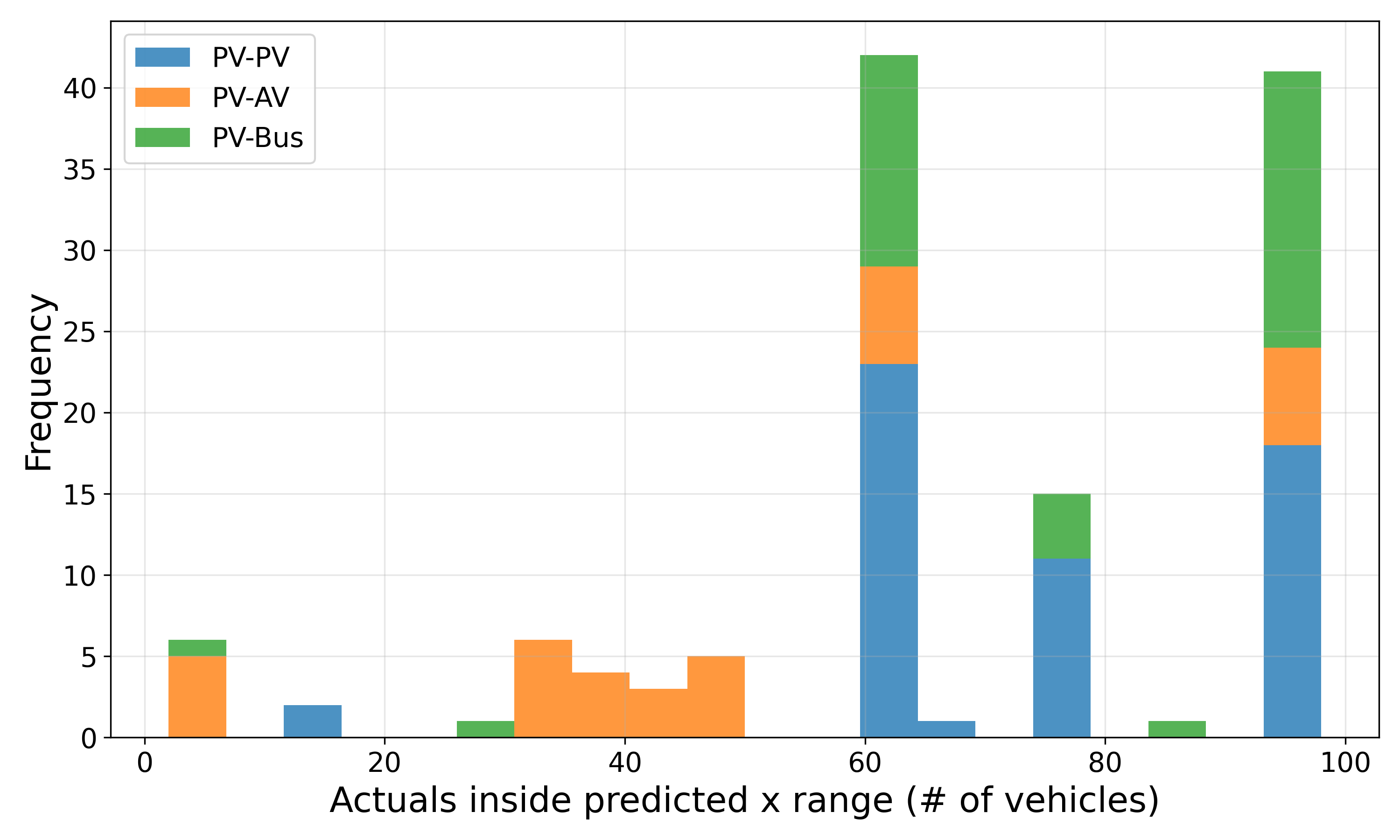}
    \caption{Stacked histogram of actual counts inside predicted $x$ range for rear-end collisions by vehicle pair.}
    \label{fig:rearend_x}
\end{figure}

\begin{figure}[t]
    \centering
    \includegraphics[width=0.8\linewidth]{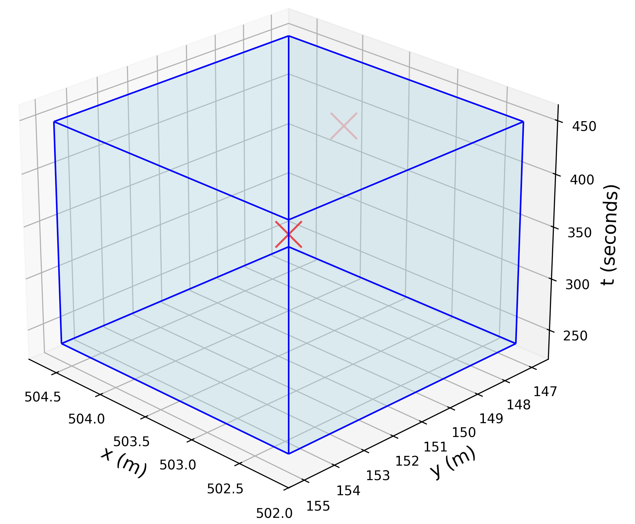}
    \caption{Predicted ranges in $(x,y,t)$ space for intersection collisions compared to the actual observed collisions, as shown by the red nodes.}
    \label{fig:intersection_alt}
\end{figure}

Figs.~\ref{fig:rearend_t}--\ref{fig:intersection_alt} present the containment performance of the BiLSTM in predicting collision time $t$ and spatial coordinates $(x,y)$. The stacked histograms report the number of actual events captured within the predicted intervals across different collision types.

For \textit{rear-end collisions}, the model consistently captures most actual events within the predicted ranges. PV-PV collisions dominate the dataset, producing high frequencies at tighter containment levels (see Figs.~\ref{fig:rearend_t}--\ref{fig:rearend_x}). Predictions for PV-AV and PV-Bus collisions are more dispersed but still achieve reasonable coverage, indicating that the model generalizes well to heterogeneous vehicle interactions. Notably, temporal containment $t$ is particularly strong, with most collisions captured within narrow time horizons. In Fig.~\ref{fig:rearend_y}, PV-Bus collisions exhibit wider dispersion along the y-axis compared to PV-PV and PV-AV cases, reflecting greater variability in how buses interact spatially with trailing vehicles. 

For \textit{intersection collisions} (PV-PV only), predictions are even sharper. As shown in Fig.~\ref{fig:intersection_alt}, the model achieves near-complete containment with narrow predicted intervals across $t$, $x$, and $y$. This performance reflects the deterministic spatiotemporal structure of intersection events and the fact that we conducted experiments on a single intersection geometry at varying times. The reduced variability in roadway context likely contributes to the higher prediction accuracy compared to rear-end collisions. As we can see, the predicted intervals tightly overlap with the ground-truth $(x,y,t)$ collision points, with only minor temporal deviations, demonstrating the model’s ability to provide reliable containment ranges suitable for real-time event flagging.

\subsubsection{Network-Level Impacts}
\begin{figure}[t]
    \centering
    \includegraphics[width=0.99\linewidth]{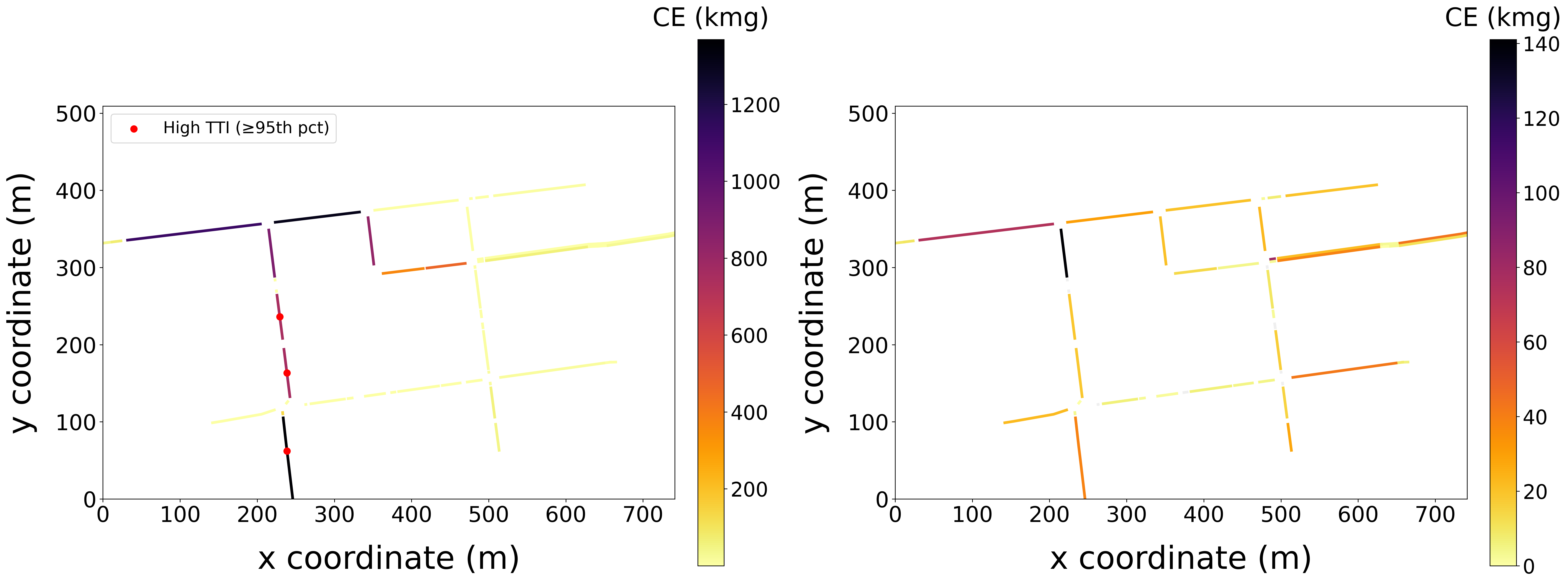}
    \caption{Total CE compared between the experimental (left) and free-flow (right) scenarios.}
    \label{fig:co2_pred}
\end{figure}

\begin{figure}[t]
    \centering
    \includegraphics[width=0.99\linewidth]{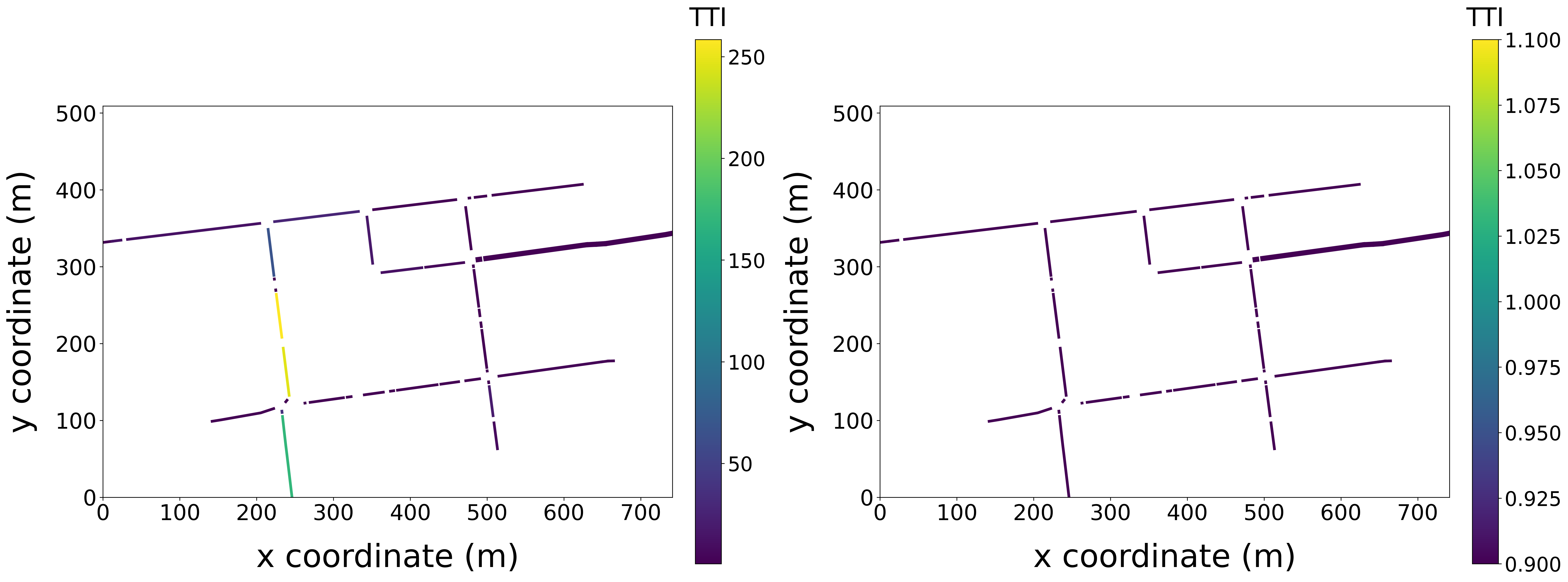}
    \caption{The TTI compared between experimental (right) and free-flow (left) scenarios.}
    \label{fig:tti_pred}
\end{figure}

Fig.~\ref{fig:co2_pred} and Fig.~\ref{fig:tti_pred} show the DCRNN’s predicted network-level CE emissions and TTI, respectively. These results highlight how the model captures the propagation of anomalies through the network and their sustainability consequences. As shown in Fig.~\ref{fig:co2_pred}, collisions and congestion significantly elevate emissions on affected links. The model accurately reproduces both the magnitude and spatial distribution of emissions, though in certain areas, predicted values overshoot the ground truth, particularly along heavily congested edges. Compared to the free-flow scenario, where emissions remain low and uniform, the experimental case exhibits sharp spikes near rear-end sites, reflecting the impact of stop-and-go traffic and idling. 

For TTI, the DCRNN predictions (see Fig.~\ref{fig:tti_pred}) effectively capture localized spikes induced by anomalies. Forecasts align with observed patterns of sharp increases in travel times downstream of collisions, though peak magnitudes are sometimes amplified. In the free-flow scenario, vehicles move smoothly across the network with low TTI values close to baseline conditions, indicating minimal delay. By contrast, the experimental case shows a sharp rise in TTI exceeding $10$, especially along central vertical and horizontal corridors. The comparison demonstrates that collisions and disruptions introduce localized bottlenecks that propagate through adjacent segments.

\subsubsection{Integrated Model Perspective}
The above results highlight the complementary strengths of the two modeling approaches. The BiLSTM provides fine-grained localization of collision events, achieving narrow interval predictions in both space and time (see Fig.~\ref{fig:intersection_alt}), which is critical for pinpointing disruptions as they occur in real time. The DCRNN, in contrast, captures how these localized anomalies propagate across the network, forecasting their effects on congestion and emissions at broader scales (see Figs.~\ref{fig:co2_pred}–\ref{fig:tti_pred}). Together, the models bridge event-level detection and system-level forecasting: the BiLSTM identifies where and when disruptions arise, while the DCRNN predicts how they spread and affect sustainability outcomes. This integration enables both precise detection and system-wide prediction, laying the foundation for anomaly-aware traffic management strategies.

\subsubsection{Limitations and Future Work} Despite the promising results, the framework has some limitations. The simulation was tested mainly on a representative set of scripted collisions in a single urban network, limiting scenario diversity and variability. While the DCRNN captured the spatial footprint of congestion and emissions, it occasionally over- or under-estimated magnitudes that were found to be spikes in the dataset, reflecting challenges in modeling nonlinear traffic-environment feedback. Additionally, the framework currently operates in an offline evaluation mode, without integration into live traffic management systems. These constraints suggest the need for further validation and refinement before deployment in real-world settings.

We will focus future work on extending the framework in several directions. Incorporating additional collision types, multimodal interactions, and more complex urban networks will improve the robustness of the models. Embedding adaptive control strategies such as real-time rerouting or signal timing within the prediction loop could demonstrate how anomaly-aware forecasts directly support mitigation and sustainability objectives. Finally, exploring online learning and real-time deployment through traffic sensor streams would enable proactive responses to emerging disruptions. By bridging predictive modeling with control and decision support, the proposed framework has the potential to advance sustainable and resilient urban mobility.  

\section{Conclusion}\label{Section VI}
In this paper, we presented a simulation-based framework that unifies controlled anomaly generation, metric logging, and hybrid deep learning to study the impacts of collisions and congestion on urban traffic networks.  We showed that by integrating BiLSTM models with a DCRNN for network-level forecasting, the framework can detect, localize, and predict anomalies while quantifying their effects on both mobility and sustainability. Our results show that the BiLSTM achieves strong containment accuracy, particularly for intersection collisions, while the DCRNN effectively translates localized disruptions into forecasts of travel time index and CE across the network. These contributions establish a reproducible and extensible pipeline for resilient traffic management research. A potential direction of future research should focus on extending the framework with multimodal interactions and larger urban networks to strengthen model robustness. In addition, embedding adaptive control strategies such as real-time rerouting will enable proactive responses to disruptions.

\bibliographystyle{IEEEtran}
\bibliography{myBib, IDS_Publications}  

\end{document}